# Mean-field treatment of the linear $\sigma$ model in dynamical calculations of DCC observables*

*Jørgen Randrup*

Nuclear Science Division, Lawrence Berkeley National Laboratory
University of California, Berkeley, California 94720

December 18, 1996

*Abstract:*

Approximate mean-field equations of motion for the classical chiral field are developed within the linear $\sigma$ model by means of a Hartree factorization. Both the approximate and the unapproximated equations of motion are augmented with a Rayleigh cooling term to emulate a uniform expansion, thereby allowing the extraction of observables relevant to the detection of disoriented chiral condensates, specifically the pion power spectrum, the pion correlation function, and the distribution of the neutral pion fraction. While the mean-field dynamics appears to be sufficiently accurate to be practically useful, the results also illustrate some difficulties associated with identifying the phenomenon experimentally.

*This work was supported by the Director, Office of Energy Research, Office of High Energy and Nuclear Physics, Nuclear Physics Division of the U.S. Department of Energy under Contract No. DE-AC03-76SF00098.

# 1 Introduction

High-energy collisions of hadrons or nuclei may produce extended regions of space within which chiral symmetry is temporarily nearly restored and the subsequent non-equilibrium relaxation towards the normal vacuum may then produce *disoriented chiral condensates*, a coherent oscillation of the pion field that is manifested through anomalous pion multiplicity distributions [1, 2, 3, 4, 5]. In order to assess the prospects for such *DCC* phenomena to actually occur and be detectable above the background, it is necessary to perform extensive dynamical calculations. Most dynamical studies have been carried out within the framework of the linear $\sigma$ model [6], with the chiral degrees of freedom being represented by a classical field [7, 8, 9, 10, 11, 12, 13, 14, 15, 16, 17, 18]. For recent reviews of the topic, see refs. [19, 20].

The purpose of this paper is two-fold. The first is to derive approximate equations of motions by means of the Hartree factorization technique, by proceeding in a manner analogous to what was done recently in the context of statistical equilibrium [21]. The resulting mean-field description is practically simpler to treat and it provides a useful conceptual framework for understanding the dynamics. Additionally, we wish to employ the linear $\sigma$ model for the calculation of key observables relevant to the detection of disoriented chiral condensates, namely the power spectrum of the emerging free pions, their correlation function, and the associated distribution of the neutral pion fraction. For this purpose, we emulate a Bjorken-type expansion in $D$ dimensions by augmenting the equations of motion with a simple Rayleigh cooling term. Starting from thermal equilibrium, we then follow the dynamical evolution of the chiral field until large times when the asymptotic regime of decoupling has been reached and the observables can be extracted. Comparisons between the results of the mean-field dynamics with those of the non-approximated equation of motion provide a basis for assessing the approximation which appears to be sufficiently accurate to be of practical utility. Moreover, the calculated observables are interesting in their own right, as they illustrate the key features of the *DCC* phenomenon and bring out some of the difficulties inherent in the experimental search for signals.

The presentation is organized as follows. In sect. 2 we develop the mean-field equations of motion and, in addition, we indicate how the approximation can be utilized for statistical equilibrium. Subsequently, in sect. 3, we describe how field configurations corresponding to various physical scenarios of interest can be prepared and treated. We then turn to the discussion of the extracted observables, sect. 4, and finally make our concluding discussion in sect. 5.

## 2 The mean-field approximation

The linear $\sigma$ model [6] offers a convenient tool for exploring the non-equilibrium dynamics of *DCC* formation. In the usual approximation, it describes the chiral degrees of freedom by means of a real classical field which forms an $O(4)$ vector, $\boldsymbol{\phi}(\boldsymbol{r}) = (\sigma(\boldsymbol{r}), \boldsymbol{\pi}(\boldsymbol{r}))$, and is subject to a non-linear interaction, $V = \frac{\lambda}{4}(\phi^2 - v^2)^2 - H\sigma$. (We use units where $\hbar$ and $c$ are unity, and $\phi^2 = \boldsymbol{\phi} \circ \boldsymbol{\phi} = \sigma^2 + \boldsymbol{\pi} \cdot \boldsymbol{\pi}$ where $\circ$ denotes the $O(4)$ scalar product.) The equation of motion for the chiral field $\boldsymbol{\phi}(\boldsymbol{r}, t)$ is then

$$[\Box + \lambda(\phi^2 - v^2)]\boldsymbol{\phi} \; = \; H\hat{\sigma} \; , \tag{1}$$

where the three parameters, $\lambda, v, H$ are adjusted to ensure reproduction of the ground-state field strength $\phi_{\rm vac} = f_\pi$ (=92 MeV) and the free meson masses $m_\pi$ (=138 MeV) and $m_\sigma$ (=600 MeV).

For the present study, we enclose the system in a cubic box with volume $\Omega = L^3$ and impose periodic boundary conditions. It is then possible to uniquely decompose the field,

$$\boldsymbol{\phi}(\boldsymbol{r}, t) \; = \; \underline{\boldsymbol{\phi}}(t) \; + \; \delta\boldsymbol{\phi}(\boldsymbol{r}, t) \; = \; \sum_{\mathbf{k}} \boldsymbol{\phi}_{\mathbf{k}}(t) \; \mathrm{e}^{i\mathbf{k}\cdot\boldsymbol{r}} \; , \tag{2}$$

$$\boldsymbol{\psi}(\boldsymbol{r}, t) \; = \; \underline{\boldsymbol{\psi}}(t) \; + \; \delta\boldsymbol{\psi}(\boldsymbol{r}, t) \; = \; \sum_{\mathbf{k}} \boldsymbol{\psi}_{\mathbf{k}}(t) \; \mathrm{e}^{i\mathbf{k}\cdot\boldsymbol{r}} \; . \tag{3}$$

The underscored part is the spatial average over the volume of the box, $\underline{\boldsymbol{\phi}} =< \boldsymbol{\phi} >$ and $\underline{\boldsymbol{\psi}} =< \boldsymbol{\psi} >$. It will be referred to as the "order parameter", while the remaining fluctuating part represents quasiparticle excitations relative to that constant background field. The coefficients in the Fourier expansions shown on the right are simply given by spatial averages, $\boldsymbol{\phi}_{\mathbf{k}} =< \boldsymbol{\phi} \exp(-i\mathbf{k}\cdot\boldsymbol{r}) >$ and $\boldsymbol{\psi}_{\mathbf{k}} =< \boldsymbol{\psi} \exp(-i\mathbf{k}\cdot\boldsymbol{r}) >$. The terms corresponding to $k = 0$ then represent the order parameter $\boldsymbol{\phi}_{\mathbf{0}} = \underline{\boldsymbol{\phi}}$ and its time derivative $\boldsymbol{\psi}_{\mathbf{0}} = \underline{\boldsymbol{\psi}}$. Moreover, since the fields are real, the Fourier coefficients satisfy symmetry relations, $\boldsymbol{\phi}^*_{\mathbf{k}} = \boldsymbol{\phi}_{-\mathbf{k}}$ and $\boldsymbol{\psi}^*_{\mathbf{k}} = \boldsymbol{\psi}_{-\mathbf{k}}$.

### 2.1 Equations of motion

By inserting the decomposition $\boldsymbol{\phi} = \underline{\boldsymbol{\phi}} + \delta\boldsymbol{\phi}$ into the equation of motion (1) and subsequently taking the spatial average, it is possible to derive an equation for the order parameter,

$$\Box\underline{\boldsymbol{\phi}} \; + \; \underline{\boldsymbol{M}} \circ \underline{\boldsymbol{\phi}} \; = \; H\hat{\sigma} \; , \tag{4}$$

where the $O(4)$ tensor $\underline{\boldsymbol{M}}$ is given by

$$\underline{\boldsymbol{M}} \; = \; \lambda(< \phi^2 > - v^2)\boldsymbol{I} \; + \; 2\lambda < \delta\boldsymbol{\phi}\delta\boldsymbol{\phi} > \; , \tag{5}$$

with $\boldsymbol{I}$ being the $O(4)$ unit tensor. We note that $< \phi^2 >= \phi_0^2 + < \delta\phi^2 >$ where $\phi_0$ is the magnitude of the order parameter, $\phi_0^2 = \underline{\boldsymbol{\phi}} \circ \underline{\boldsymbol{\phi}}$. The equation (4) was first derived by Baym and Grinstein [22]. Although exact, it is of little practical value unless the averages entering in the mass tensor (5) are provided.

Subtracting now the equation (4) for the average field strength from the full equation of motion (1), we obtain an approximate equation of motion for the field fluctuations. It can be written on a simple Klein-Gordon form,

$$\Box\delta\boldsymbol{\phi} \;+\; \boldsymbol{M}\circ\delta\boldsymbol{\phi} \;\approx\; \mathbf{0}\;, \tag{6}$$

where the quasi-particle mass tensor $\boldsymbol{M}$ is

$$\boldsymbol{M} \;=\; \lambda(<\phi^2> - v^2)\boldsymbol{I} \;+\; 2\lambda <\boldsymbol{\phi}\boldsymbol{\phi}> = \; \underline{\boldsymbol{M}} \;+\; 2\lambda\underline{\boldsymbol{\phi}\boldsymbol{\phi}}\;. \tag{7}$$

The last relation follows because $<\boldsymbol{\phi}\boldsymbol{\phi}> = \underline{\boldsymbol{\phi}\boldsymbol{\phi}} + <\delta\boldsymbol{\phi}\delta\boldsymbol{\phi}>$. In deriving this result, we have employed a Hartree linearization. Using $j$=0,1,2,3 to denote the $O(4)$ components of the chiral field, the approximation can be characterized as follows: Terms containing an even number of fluctuations are replaced by their spatial average, $\delta\phi_j\delta\phi_{j'} \rightarrow <\delta\phi_j\delta\phi_{j'}>$ (which leads to a cancellation of the even terms) and the odd fluctuation terms are contracted,

$$\delta\phi_j\delta\phi_{j'}\delta\phi_{j''} \;\rightarrow <\delta\phi_j\delta\phi_{j'}>\delta\phi_{j''} \;+\; <\delta\phi_{j'}\delta\phi_{j''}>\delta\phi_j \;+\; <\delta\phi_j''\delta\phi_j>\delta\phi_{j'}\;. \tag{8}$$

Generally, one could imagine an effective theory containing higher powers of the field (although it could not have arisen from a renormalizable quantum-field theory). The linearization method can be applied in this more complicated case as well by simply employing the gaussian approximation to reduce higher-order terms.

Since the present treatment defines the order parameter $\underline{\boldsymbol{\phi}}$ as the average of the field amplitude over the entire volume of the box, it has no spatial dependence and the d'Alambert operator $\Box$ then reduces to $\partial_t^2$. However, it is possible to generalize the definition of the order parameter so that it does vary in space (see sect. 3.2). The above mean-field equations can then still be employed and with a view towards this more realistic situation we leave the full d'Alambert operator in place.

The above developments are in accordance with earlier work by Boyanovsky *et al.* [13]. However, in so far as the dimensionality of the chiral space can be regarded as large, $N \gg 1$, its inverse provides a useful expansion parameter, and it may then be natural to keep only the leading terms in $1/N$ [7, 8, 12, 13, 16, 17, 23, 24]. The term $2\lambda <\delta\boldsymbol{\phi}\delta\boldsymbol{\phi}>$ is then absent from the mass tensors $\underline{\boldsymbol{M}}$ and $\boldsymbol{M}$ given in eqs. (5) and (7). This term is of relative order $2/N$, since the leading term in the effective masses contains $N$ individual fluctuation terms $<\delta\phi_j^2>$, and there are no higher-order terms. Since the actual dimensionality is not all that large, $N$=4, the inclusion of the $1/N$ terms produces a 50% increase in the contribution from the quasi-particle degrees of freedom to the effective masses. Changes in the fluctuations of such relatively large magnitude may have a significant influence on the calculated results and we therefore prefer to keep all the terms. (It should be noted that the inclusion of the term $2\lambda <\delta\boldsymbol{\phi}\delta\boldsymbol{\phi}>$ has no impact on the computational burden, since it merely modifies the magnitude of the appropriate coefficient.) As we shall see, the mean-field treatment with the $1/N$ term retained provides a practically useful approximation to the full equation of motion (1).

### 2.1.1 k representation

The mean-field description derived above provides a very convenient means for obtaining approximate solutions for the dynamical evolution of the chiral field. When reexpressed in the $k$ representation by means of a Fourier transformation, the coupled equations of motion are as follows,

$$k = 0 \quad : \qquad \partial_t^2 \boldsymbol{\phi}_{\mathbf{0}} \; + \; \underline{\boldsymbol{M}} \circ \boldsymbol{\phi}_{\mathbf{0}} \; = \; H\hat{\sigma} \; , \tag{9}$$

$$k > 0 \quad : \qquad (\partial_t^2 \; + \; k^2)\boldsymbol{\phi}_{\mathbf{k}} \; + \; \boldsymbol{M} \circ \boldsymbol{\phi}_{\mathbf{k}} \; = \; \mathbf{0} \; , \tag{10}$$

where the mass tensors are given by

$$\boldsymbol{M} \; = \; \lambda(\sum_{\mathbf{k}} \boldsymbol{\phi}_{\mathbf{k}}^* \circ \boldsymbol{\phi}_{\mathbf{k}} \; - \; v^2)\boldsymbol{I} \; + \; 2\lambda \sum_{\mathbf{k}} \boldsymbol{\phi}_{\mathbf{k}} \boldsymbol{\phi}_{\mathbf{k}}^* \; = \; \underline{\boldsymbol{M}} \; + \; 2\lambda \; \boldsymbol{\phi}_{\mathbf{0}} \boldsymbol{\phi}_{\mathbf{0}} \; . \tag{11}$$

The above equations (9-11) form a simple self-contained system that can be solved at only modest numerical effort. In sect. 4 we shall illustrate the utility of this convenient approximate description.

## 2.2 Energy

The total energy associated with a given dynamical state of the chiral field can be written as $\Omega E[\boldsymbol{\psi}, \boldsymbol{\phi}]$, where $E[\boldsymbol{\psi}, \boldsymbol{\phi}]$ is the average energy density. This key quantity can be decomposed into terms having instructive physical interpretations, thereby enabling us to develop a simple approximate treatment of statistical equilibrium [21],

$$E[\boldsymbol{\psi}, \boldsymbol{\phi}] \; = < \frac{1}{2}\psi^2 + \frac{1}{2}(\nabla\boldsymbol{\phi})^2 + \frac{\lambda}{4}(\phi^2 - v^2)^2 - H\sigma > = \; E_0 \; + \; E_{\rm qp} \; + \; \delta V \; . \tag{12}$$

Here $E_0$ is the energy density arising if all the field fluctuations were put to zero,

$$E_0 \; \equiv \; E[\boldsymbol{\psi} = \underline{\boldsymbol{\psi}}(\boldsymbol{r}), \boldsymbol{\phi} = \underline{\boldsymbol{\phi}}(\boldsymbol{r})] \; = \; \frac{1}{2}\psi_0^2 \; + \; \frac{\lambda}{4}(\phi_0^2 - v^2)^2 \; - \; H\phi_0 \cos\chi_0 \; , \tag{13}$$

where $\psi_0^2 = \underline{\boldsymbol{\psi}} \circ \underline{\boldsymbol{\psi}}$ and the disorientation angle $\chi_0$ measures the angle between the order parameter $\underline{\boldsymbol{\phi}}$ and the vacuum direction $\hat{\sigma}$, $\cos\chi_0 = \hat{\underline{\boldsymbol{\phi}}} \circ \hat{\sigma}$. The first term in $E_0$ is the bare kinetic energy density $K_0$ and the remainder is the bare interaction energy density $V_0$.

The second term in (12) is given by

$$E_{\rm qp}[\boldsymbol{\psi}, \boldsymbol{\phi}] \; = \; \frac{1}{2} < \delta\boldsymbol{\psi} \circ \delta\boldsymbol{\psi} \; + \; \nabla\delta\boldsymbol{\phi} \circ \nabla\delta\boldsymbol{\phi} \; + \; \delta\boldsymbol{\phi} \circ \boldsymbol{M} \circ \delta\boldsymbol{\phi} > \; , \tag{14}$$

where $\boldsymbol{M}$ is the quasiparticle mass tensor (7) derived in connection with the mean-field equations of motion above. Therefore $\Omega E_{\rm qp}$ can be interpreted as the energy associated with the gas of independent quasiparticles described by the Klein-Gordon equation of motion (6).

The last term in (12) corrects for the fact that the interaction is non-linear. It can be written as

$$\delta V[\boldsymbol{\psi},\boldsymbol{\phi}] \;=\; \frac{\lambda}{4} < \delta\phi^4 > \;-\; \frac{\lambda}{2} < \delta\phi^4 >_G \;+\; \lambda\underline{\boldsymbol{\phi}} \circ < \delta\boldsymbol{\phi}\;\delta\phi^2 > \;, \qquad (15)$$

where $< \cdot >_G$ denotes the evaluation of the average by means of the gaussian approximation,

$$\begin{aligned} < \delta\phi^4 >_G \;&=\; \sum_{jj'} [< \delta\phi_j\delta\phi_j >< \delta\phi_{j'}\delta\phi_{j'} > \;+\; 2 < \delta\phi_j\delta\phi_{j'} >< \delta\phi_{j'}\delta\phi_j >] \\ &=\; < \delta\phi^2 >^2 \;+\; 2\;\mathrm{tr}(< \delta\boldsymbol{\phi}\delta\boldsymbol{\phi} > \circ < \delta\boldsymbol{\phi}\delta\boldsymbol{\phi} >)\;. \end{aligned} \qquad (16)$$

It is important to recognize that the above relations are *exact*, since they are merely reformulations of the expression (12) for the energy of a given state.

## 2.3 Statistical equilibrium

We now briefly indicate how the mean-field approximation can be applied to statistical equilibrium, leading to the simple approximate treatment developed in ref. [21].

The natural starting point is the partition function,

$$\mathcal{Z}_T \;=\; \int \mathcal{D}[\boldsymbol{\psi},\boldsymbol{\phi}]\; \mathrm{e}^{-\frac{\Omega}{T}E[\boldsymbol{\psi},\boldsymbol{\phi}]}\;, \qquad (17)$$

where the functional integral is over all physically possible configurations of the chiral field $(\boldsymbol{\psi}(\boldsymbol{r}),\boldsymbol{\phi}(\boldsymbol{r}))$. For the purpose of treating the statistical equilibrium, we now make a number of approximations. The first and most important approximation is to replace the state-dependent correction term $\delta V$ (15) by its thermal average, for each given value of the order parameter $(\underline{\boldsymbol{\psi}},\underline{\boldsymbol{\phi}})$. This is a good approximation when the system considered extends over a volume that is large on the scale of the correlation length, since the fluctuations in one part of the system are then independent of those in other parts and so the spatial average $< \cdot >$ approaches the corresponding thermal average $\prec \cdot \succ$.

The last term in (15) then vanishes by symmetry, since it contains an odd number of field fluctuations. Moreover, we employ the gaussian approximation (16) for the evaluation of the first term in (15). We can simplify the analysis by taking advantage of the fact that, for any given value of the order parameter $\underline{\boldsymbol{\phi}}$, the tensor $\prec \delta\boldsymbol{\phi}\delta\boldsymbol{\phi} \succ$ is aligned with the order parameter and, furthermore, is invariant under rotations around that direction. It then follows that

$$\mathrm{tr}(\prec \delta\boldsymbol{\phi}\delta\boldsymbol{\phi} \succ) \;=\; \prec \delta\phi^2 \succ \;=\; \prec \delta\phi_\parallel^2 \succ \;+\; 3 \prec \delta\phi_\perp^2 \succ\;, \qquad (18)$$

$$\mathrm{tr}(\prec \delta\boldsymbol{\phi}\delta\boldsymbol{\phi} \succ \circ \prec \delta\boldsymbol{\phi}\delta\boldsymbol{\phi} \succ) \;=\; \prec \delta\phi_\parallel^2 \succ^2 \;+\; 3 \prec \delta\phi_\perp^2 \succ^2\;, \qquad (19)$$

where $\delta\phi_\parallel$ denotes the component of $\delta\boldsymbol{\phi}$ in the direction of the order parameter, $\delta\phi_\parallel = \delta\boldsymbol{\phi} \circ \hat{\underline{\boldsymbol{\phi}}}$ and $\prec \delta\phi_\perp^2 \succ$ denotes the variance of the thermal field fluctuations in

(any) one of the perpendicular directions. These approximations amount to a simple replacement,

$$\delta V[\boldsymbol{\psi},\boldsymbol{\phi}] \;\to\; \delta\tilde{V}(\underline{\boldsymbol{\psi}},\underline{\boldsymbol{\phi}}) \;\equiv\; -\frac{\lambda}{4} \prec \delta\phi^4 \succ_G = \tag{20}$$

$$-\frac{\lambda}{4}\left[3 \prec \delta\phi_\parallel^2 \succ^2 \; + 6 \prec \delta\phi_\parallel^2 \succ\prec \delta\phi_\perp^2 \succ \; + 15 \prec \delta\phi_\perp^2 \succ^2\right] . \tag{21}$$

Once $\delta V$ in the energy expression (12) has been approximated in this manner, the partition function can be brought onto a convenient factorized form,

$$\mathcal{Z}_T \;\approx\; \int d^4\underline{\boldsymbol{\psi}}d^4\underline{\boldsymbol{\phi}}\; \mathrm{e}^{-\frac{\Omega}{T}(E_0+\delta\tilde{V})} \prod_{\mathbf{k}}{}' \left[\int d^4\boldsymbol{\psi}_{\mathbf{k}}d^4\boldsymbol{\phi}_{\mathbf{k}}\right]\; \mathrm{e}^{-\frac{\Omega}{T}E_{\rm qp}} \;, \tag{22}$$

where the functional integral over the states of the chiral field has been expressed in terms of the Fourier components introduced in eqs. (2-3). The prime indicates the omission of the zero mode having $\mathbf{k} = \mathbf{0}$. Since the state-dependent quasiparticle energy $E_{\rm qp}$ is a sum of contributions from each wave number $\mathbf{k}$,

$$E_{\rm qp} \;=\; \frac{1}{2}\sum_{\mathbf{k}}{}'[\boldsymbol{\psi}_{\mathbf{k}}\circ\boldsymbol{\psi}_{\mathbf{k}} \;+\; k^2\boldsymbol{\phi}_{\mathbf{k}}\circ\boldsymbol{\phi}_{\mathbf{k}} \;+\; \boldsymbol{\phi}_{\mathbf{k}}\circ\boldsymbol{M}\circ\boldsymbol{\phi}_{\mathbf{k}}] \;, \tag{23}$$

the multiple integral over the fluctuations factorizes correspondingly.

The expression for the partition function may then be reduced to an integral over the order parameter (and its time derivative),

$$\mathcal{Z}_T \;=\; \int d^4\underline{\boldsymbol{\psi}}d^4\underline{\boldsymbol{\phi}}\; W_T(\underline{\boldsymbol{\psi}},\underline{\boldsymbol{\phi}}) \;, \tag{24}$$

with the associated statistical weight being given by

$$W_T(\underline{\boldsymbol{\psi}},\underline{\boldsymbol{\phi}}) \;\approx\; \mathrm{e}^{-\frac{\Omega}{T}(E_0+\delta\tilde{V}+\prec E_{\rm qp}\succ-TS_T)} \;=\; \mathrm{e}^{-\frac{\Omega}{T}(K_0+V_T-TS_T)} \;=\; \mathrm{e}^{-\frac{\Omega}{T}(K_0+F_T)} \;. \tag{25}$$

The last two terms in the first exponent arise from the integration over the quasiparticle degrees of freedom in eq. (22). They represent the density of energy and entropy of the quasiparticle gas,

$$\prec E_{\rm qp}\succ \;=\; \frac{1}{\Omega}\sum_{\mathbf{k}}{}'\left[\epsilon_k^\parallel f_k^\parallel \;+\; 3\epsilon_k^\perp f_k^\perp\right] \;, \tag{26}$$

$$S_T \;=\; \frac{1}{\Omega}\sum_{\mathbf{k}}{}'\left[\bar{f}_k^\parallel \ln\bar{f}_k^\parallel - f_k^\parallel\ln f_k^\parallel \;+\; 3(\bar{f}_k^\perp\ln\bar{f}_k^\perp - f_k^\perp\ln f_k^\perp)\right] \;. \tag{27}$$

The thermal occupation probabilities are of Bose-Einstein form,[1] $f_k^\parallel = (\mathrm{e}^{\epsilon_k^\parallel/T}-1)^{-1}$, with $\bar{f}_k^\parallel = 1+f_k^\parallel$, and similarly for $f_k^\perp$. Furthermore, $\epsilon_k^\parallel$ and $\epsilon_k^\perp$ denote the frequencies of a given quasiparticle mode, $\epsilon_k^2 = k^2 + \mu^2$, where $\mu^2$ is the appropriate eigenvalue of the quasiparticle mass tensor $\boldsymbol{M}$,

$$\mu_{\parallel}^2 = \lambda(3\phi_0^2 + 3 \prec \delta\phi_{\parallel}^2 \succ + 3 \prec \delta\phi_{\perp}^2 \succ - v^2) , \tag{28}$$
$$\mu_{\perp}^2 = \lambda(\phi_0^2 + \prec \delta\phi_{\parallel}^2 \succ + 5 \prec \delta\phi_{\perp}^2 \succ - v^2) . \tag{29}$$

[1] This result appears when the field configurations considered are restricted to those for which any given quasiparticle mode contains an integral number of elementary quanta, each having the energy $\epsilon_k$, thereby reducing the functional integral over the fluctuations to a multiple sum over occupation numbers $n_{\mathbf{k}}^{(j)}$. This condition is appropriate in the present context where the underlying system is of quantal nature; treating the occupation numbers as continuous variables would lead to a Maxwell-Boltzmann form, $f_k = \exp(-\epsilon_k/T)$.

In these expressions the fluctuations are given self-consistently,

$$\prec \delta\phi_{\parallel}^2 \succ = \sum_{\mathbf{k}}{}' \frac{f_k^{\parallel}}{\epsilon_k^{\parallel}} = \sum_{\mathbf{k}}{}' \frac{1}{\epsilon_k^{\parallel}} \frac{1}{\mathrm{e}^{\epsilon_k^{\parallel}/T} - 1} , \tag{30}$$
$$\prec \delta\phi_{\perp}^2 \succ = \sum_{\mathbf{k}}{}' \frac{f_k^{\perp}}{\epsilon_k^{\perp}} = \sum_{\mathbf{k}}{}' \frac{1}{\epsilon_k^{\perp}} \frac{1}{\mathrm{e}^{\epsilon_k^{\perp}/T} - 1} . \tag{31}$$

In the last two expressions in eq. (25), the exponent has been rearranged for instructive purposes. The quantity $K_0(\psi_0) = \frac{1}{2}\psi_0^2$ is the bare kinetic energy density associated with the evolution of the order parameter. Furthermore,

$$V_T(\phi_0, \chi_0) \equiv V_0 + \delta\tilde{V} + \prec E_{\mathrm{qp}} \succ , \tag{32}$$

can be regarded as an effective potential for the order parameter. Finally, the entropy density $S_T(\phi_0)$ expresses the effective number of accessible quasiparticle states for a given magnitude of the order parameter (at the specified temperature $T$), and the free energy density associated with a given order parameter $\underline{\boldsymbol{\phi}}$ is then $F_T = V_T - TS_T$. Combining the above results, we note that this latter quantity is given by

$$\begin{aligned} F_T(\underline{\boldsymbol{\phi}}) &= \frac{\lambda}{4}(\phi_0^2 - v^2)^2 - H\phi_0 \cos\chi_0 \\ &- \frac{3}{4}\lambda \left[\prec \delta\phi_{\parallel}^2 \succ^2 + 2 \prec \delta\phi_{\parallel}^2 \succ\prec \delta\phi_{\perp}^2 \succ + 5 \prec \delta\phi_{\perp}^2 \succ^2\right] \\ &+ \frac{T}{\Omega} \sum_{\mathbf{k}}{}' [\ln(1 - \mathrm{e}^{-\epsilon_k^{\parallel}/T}) + 3\ln(1 - \mathrm{e}^{-\epsilon_k^{\perp}/T})] . \end{aligned} \tag{33}$$

It is then elementary to calculate the statistical weight (25) as a function of the order parameter, yielding the function $W_T(\psi_0, \phi_0, \chi_0)$.

# 3 Dynamical scenarios

We describe in this section how field configurations corresponding to various instructive physical scenarios can be prepared and treated within the linear $\sigma$ model. As we have above, we assume throughout that the system is enclosed in a three-dimensional torus, *i.e.* a rectangular box with periodic boundary conditions.

## 3.1 Sampling

We first briefly summarize how it is possible to utilize the above mean-field treatment of equilibrium to devise a simple approximate method for preparing chiral field configurations that represent chiral matter in statistical equilibrium [21].

The first task is to sample the order parameter $(\underline{\psi}, \underline{\phi})$ on the basis of the statistical weight $W_T(\psi_0, \phi_0, \chi_0)$ given in eq. (25). This quantity factorizes, due to the additive form of the exponent. The time derivative $\psi_0$ is then governed by a four-dimensional normal distribution, $P_\psi(\underline{\psi}) \sim \exp(-\Omega\psi_0^2/2T)$. Furthermore, since the distribution of the magnitude $\phi_0$ can be pretabulated with the aid of the explicit expression (33) (ignoring the $H$ term), the associated sampling task is computationally simple. Once $\phi_0$ has been picked, the disorientation angle $\chi_0$ is easy to sample from its conditional distribution, $P_\chi(\chi_0) \sim \exp(-H\phi_0 \cos\chi_0)$, and the $O(3)$ direction $(\vartheta_0, \varphi_0)$ of $\underline{\boldsymbol{\pi}}$ is uniform on $4\pi$.

Once the magnitude of the order parameter is known, the thermal quasiparticle distributions are fully determined and the number of quanta in each elementary mode is readily sampled, using $P(n_{\mathbf{k}}) \sim \exp(-n_{\mathbf{k}}\epsilon_k/T)$ for each of the four principal chiral directions. Since the quasi-particle mass tensor is aligned with the $O(4)$ direction of the order parameter, $(\chi_0, \vartheta_0, \varphi_0)$, a subsequent $O(4)$ rotation of $\boldsymbol{\phi}(\boldsymbol{r})$ and $\boldsymbol{\psi}(\boldsymbol{r})$ must then be performed in order to express the field in the usual $(\sigma, \boldsymbol{\pi})$ reference system.

By employing this simple sampling method, it is thus possible to sample field configurations corresponding to macroscopically uniform matter at a given temperature.

## 3.2 Finite sources

The above method is appropriate for the study of uniform matter. When more realistic scenarios are considered, the hot system of physical interest is embedded in the normal vacuum in which the order parameter is $\boldsymbol{\phi}_{\rm vac} = (f_\pi, \mathbf{0})$ and $\boldsymbol{\psi}_{\rm vac} = (0, \mathbf{0})$. Such systems can be equally easily prepared by applying a simple modulation of the sampled fields.

Let the shape of the hot part of the system, the "source", be described by the modulation function $g(\boldsymbol{r})$ which drops from unity well inside the source to zero outside. For the simple case of a spherical source of radius $R$ we may use

$$g_{\rm sphere}(\boldsymbol{r}) \;=\; \frac{1}{{\rm e}^{(r-R)/a}+1}\;, \tag{34}$$

where $a$ governs the surface width. More complicated source configurations can be described similarly.

To obtain a suitable configuration $(\boldsymbol{\psi}(\boldsymbol{r}), \boldsymbol{\phi}(\boldsymbol{r}))$ describing the finite source, we first pick a field configuration $(\boldsymbol{\psi}'(\boldsymbol{r}), \boldsymbol{\phi}'(\boldsymbol{r}))$ corresponding to uniform matter at the desired temperature, for example by using the method described above. This field is then subjected to a simple spatial modulation,

$$\boldsymbol{\phi}(\boldsymbol{r}) \;=\; \boldsymbol{\phi}_{\rm vac} \;+\; g(\boldsymbol{r})[\boldsymbol{\phi}'(\boldsymbol{r}) - \boldsymbol{\phi}_{\rm vac}]\;, \tag{35}$$

$$\boldsymbol{\psi}(\boldsymbol{r}) \;=\; \boldsymbol{\psi}_{\rm vac} \;+\; g(\boldsymbol{r})[\boldsymbol{\psi}'(\boldsymbol{r}) - \boldsymbol{\psi}_{\rm vac}]\;. \tag{36}$$

The resulting a field configuration now describes hot matter at the specified temperature when analyzed inside the source region and it smoothly approaches the vacuum value outside. It can then be used as an initial condition for either the full equation of motion (1) or its mean-field approximation (4-7).

### 3.3 Pseudo-expansion

In order for disoriented chiral condensates to appear, it is necessary to allow the system to undergo a non-equilibrium relaxation towards the vacuum configuration. In high-energy collisions, such a development is expected to occur as a consequence of the cooling generated by expansion and radiation.

Various expansion scenarios have already been considered in the literature. Wang and Huang [9] sought to emulate the standard longitudinal Bjorken expansion by employing a boost-invariant grid in the beam direction while ignoring the transverse dimensions. The natural variables are then the local proper time, $\tau = \sqrt{t^2 - z^2}$, and the spatial analog of the rapidity, $\eta = \frac{1}{2}\ln[(t+z)/(t-z)]$; the evolution of the field, $\boldsymbol{\phi}(\eta, \tau)$, was obtained by solving the correspondingly transformed equation of motion using random initial conditions for the field strength at each grid point. Later on, the treatment of the longitudinal expansion scenario was modified by Asakawa *et al.* [14] who included the transverse variables $(x, y)$, while requiring the field to be strictly boost invariant, $\boldsymbol{\phi}(x, y, \tau)$. A more complete treatment of the longitudinal expansion scenario was developed by Cooper *et al.* [12] who took full account of the transverse degrees of freedom in addition to the Bjorken expansion, thus obtaining $\boldsymbol{\phi}(x, y, \eta, \tau)$. Analogous expansions in three dimensions have also been considered. Gavin and Müller [10] derived an approximate equation of motion for the order parameter in such a scenario by employing a simple time dependence of the decaying field fluctuations. A full inclusion of the expanding three-dimensional geometry was made recently by Lampert *et al.* [25], to leading order in the $1/N$ expansion, yielding $\boldsymbol{\phi}(\eta, \theta, \phi, \tau)$ with $\eta = \frac{1}{2}\ln[(t+r)/(t-r)]$ being the radial rapidity.

Since realistic expansion scenarios are both computationally cumbersome and physically intransparent, we employ in the present study the simple pseudo-expansion method suggested in ref. [18]. It consists of augmenting the equation of motion (1) by a Rayleigh dissipation term of the form $-(D/t)\partial_t\boldsymbol{\phi}$ (or $-D\boldsymbol{\psi}_{\mathbf{k}}/t$ in the $k$ representation) and starting the dynamical evolution at $t_0 = 1$ fm/$c$ (as is commonly done). This term acts as a cooling agency by causing the field fluctuations to decay in the course of time. The associated decrease of the energy density is given by $\dot{E} = -(D/t) < \psi^2 >$. Moreover, at sufficiently large times, the quasiparticle amplitudes fall off as $\sim t^{-D/2}$, as can easily be seen by exploiting the approximate decoupling that occurs when the field fluctuations become small. Consequently, the quasiparticle number density (and the associated power density) decreases as $\sim t^{-D}$ as $t \to \infty$, as is characteristic of an expansion in $D$ dimensions. In order to understand that the cooling term can be regarded as generating a pseudo-expansion, it is helpful to note that temporal part of the d'Alambert operator, $\partial_t^2$, transforms into $\partial_\tau^2 + (D/\tau)\partial_\tau$ for a Bjorken-type expansion in $D$ dimensions. Finally we note that the Rayleigh cooling term is unaffected by the Hartree factorization and so the pseudo-expansion can be included into the mean-field dynamics in the same manner.

# 4 Observables

After the above developments, we now turn to the calculation of observables relevant to the experimental detection of disoriented chiral condensates. We concentrate on the pion power spectrum, the pion correlation function, and the distribution of the neutral pion fraction which are those that have received the largest attention so far.

We consider a cubic box with a side length $L = 5 - 10$ fm and sample the initial field configurations from a thermal ensemble at the temperature $T_0 = 400$ MeV, as briefly described in sect. 3.1. The $L$ values are of typical nuclear magnitude and so the considered system may be regarded as roughly approximating the interior of the excited region formed in a high-energy collision. Moreover, based on the studies in ref. [18], the initial temperature $T_0$ is chosen so as to enhance the chance for the system to enter the unstable region; lower values will lead to more stable evolutions.

After its preparation, the system is subjected to a pseudo-expansion in $D$ dimensions. The associated attenuation of the field fluctuations then causes the system to experience a non-equilibrium relaxation towards the normal vacuum. When the fluctuations have subsided sufficiently, and the order parameter has come close to its vacuum value, their further exponential evolution can be obtained analytically. It is then straightforward to extrapolate to $t \to \infty$ and thus extract observable quantities.

## 4.1 Power spectrum

We consider first the power distribution of the pions. Generally, invoking the developed mean-field formulation, the total power density at any point in time during the dynamical evolution is simple to express,

$$P_{\rm total} \;=\; \frac{1}{\Omega}\sum_{\mathbf{k}} \frac{1}{2}[\boldsymbol{\psi}^*_{\mathbf{k}} \circ \boldsymbol{\psi}_{\mathbf{k}} \;+\; k^2\; \boldsymbol{\phi}^*_{\mathbf{k}} \circ \boldsymbol{\phi}_{\mathbf{k}} \;+\; \boldsymbol{\phi}^*_{\mathbf{k}} \circ \boldsymbol{M} \circ \boldsymbol{\phi}_{\mathbf{k}}] \;, \tag{37}$$

where $\boldsymbol{M}$ should be replaced by $\underline{\boldsymbol{M}}$ for the term with $k = 0$. From an observational perspective the interest is focussed on the asymptotic stage of the evolution, $t \to \infty$. At that late stage, the field deviates only little from its vacuum value and the different quasiparticle modes are effectively decoupled. For the pion field we then have

$$\boldsymbol{\pi}(\boldsymbol{r}, t) \;=\; \sum_{\mathbf{k}} \boldsymbol{\pi}_{\mathbf{k}}\; \mathrm{e}^{i\mathbf{k}\cdot\boldsymbol{r} - i\omega_k t} \;, \tag{38}$$

where the frequency is given by the free pion dispersion relation, $\omega_k^2 = k^2 + m_\pi^2$. The spectral distribution of the power density for the emerging free pions is then given by

$$\frac{dP_\pi}{dE} \;=\; \frac{1}{\Omega}\sum_{\mathbf{k}} \omega_k^2\; \boldsymbol{\pi}^*_{\mathbf{k}} \cdot \boldsymbol{\pi}_{\mathbf{k}}\; \delta(\omega_k - m_\pi - E) \;, \tag{39}$$

where $E$ is the specified kinetic energy of the observed pion. In thermal equilibrium the corresponding quantity would be the power density associated with the modes perpendicular to the order parameter,

$$\frac{dP_\perp}{dE} \;=\; \frac{1}{\Omega}\; \sum_{\mathbf{k}} \epsilon_k^\perp\; f_k\; \delta(\epsilon_k - \mu_\perp - E) \;, \tag{40}$$

relying again on the mean-field approximation.

The results are summarized in fig. 1. The upper-left panel shows the equilibrium power density (40) for the initial temperature of $T_0 = 400$ MeV. The solid histogram is for the employed side length of $L = 5$ fm, while the dashed histogram is for an eight times larger cube, $L = 10$ fm. The slight dependence on the magnitude of the volume is primarily due to the shell effects associated with the smaller cube.

The remaining three panels in fig. 1 show the power spectra (39) resulting from pseudo-expansions in various dimensions, $D$=1,2,3. For each value of $D$, 100 initial field configurations have been sampled from the thermal ensemble having a temperature of $T_0 = 400$ MeV. As already observed in ref. [18], for relatively slow cooling rates, $D \lesssim 1$, the evolutions are nearly adiabatic and so there is little effect on the power spectrum. When the dimensionality of the expansion is sufficiently large, $D \gtrsim 2$, the system is effectively quenched into an unstable configuration. The low-energy pion modes then experience an exponential growth leading to a corresponding enhancement in the power spectrum [7, 8, 12, 13, 14]. (This low-energy enhancement causes an overall reduction elsewhere, since the displayed spectra have been given a common normalization, for convenience.) The resulting distortion of the power spectrum has been advocated as a possible signal for *DCC* formation [26].

The results obtained for $D$=1,2,3 with the mean-field dynamics, eqs. (9-11), are included as the dashed histograms in fig. 1. Generally, these results are quite similar to those obtained with the unapproximated equation of motion (1) and so the dynamical distortion of the relaxing chiral field appears to be well approximated by the mean-field equations.

## 4.2 Correlation function

It is particularly interesting to calculate the correlation function of the pion field, since this quantity determines the spectral distribution of the emerging pions [21, 26, 27, 28]. We shall therefore extract the equal-time pion correlation function at large times when there is no longer any time dependence, except for the overall decay caused by the cooling term,

$$C_\pi(r_{12}) \equiv < \delta\boldsymbol{\pi}(\boldsymbol{r}+\boldsymbol{r}_{12},t)\cdot\delta\boldsymbol{\pi}(\boldsymbol{r},t) > / < \delta\boldsymbol{\pi}(\boldsymbol{r},t)\cdot\delta\boldsymbol{\pi}(\boldsymbol{r},t) > \quad . \tag{41}$$

Here $< \cdot >$ indicates a spatial average over the position $\boldsymbol{r}$ within the volume of the torus, $\Omega$, as well as a subsequent average over the ensemble of histories considered. The normalization is such that the correlation function is unity at zero separation, $C_\pi(0) = 0$. The translational symmetry of the scenario implies that the spatial dependence is via the separation $\boldsymbol{r}_{12} = \boldsymbol{r}_1 - \boldsymbol{r}_2$ only. Moreover, to the extent that there is invariance under spatial rotations, only the magnitude $r_{12} = |\boldsymbol{r}_{12}|$ enters. In principle, the rotational symmetry is broken by the cartesian lattice, but this effect is relatively unimportant.

For each field configuration considered, the correlation function has been extracted in two different ways. One way is to calculate $C_\pi^{xyz} \equiv (C_\pi^x + C_\pi^y + C_\pi^z)/3$, where $C_\pi^x(r_{12})$ has been obtained by assuming that the relative separation $\boldsymbol{r}_{12}$ is aligned with the $x$ axis, and analogously for $C_\pi^y$ and $C_\pi^z$. The other way yields $C_\pi^{\rm diag}$ which is extracted by aligning $\boldsymbol{r}_{12}$ with the (111) diagonal. For a cube with side length $L$, $C_\pi^{xyz}$ has a periodicity of $L$ (thus effectively limiting the separation $r_{12}$ to $L/2$), while $C_\pi^{\rm diag}$ has a periodicity that is $\sqrt{3}$ times larger, thereby making better use of the limited volume of the lattice. The difference between the two extracted correlation functions may serve as an indication of the combined effect of the incomplete rotational symmetry arising from the cartesian lattice and the finite number of histories considered.

The resulting correlation function for the transverse quasi-particle modes in thermal equilibrium is shown in the upper-left panel of fig. 2, for a number of box sizes, $L = 5, 8, 10$ fm and using a temperature of $T_0 = 400$ MeV. The correlation function is neither sensitive to the box size nor to the particular method of extraction, thus giving confidence in the method employed. Moreover, it is in good agreement with the analytical result for infinite matter (see ref. [21]).

The three other panels of fig. 2 show the correlation functions extracted at large times from the same systems as those considered for the above calculation of the power spectrum. Since, roughly speaking, the correlation function and the power spectrum are related by a Fourier transformation, the preceding discussion already suggests what behavior to expect. Thus, the correlation function for $D = 1$ is rather similar to the thermal form, in accordance with the small distortion exhibited by the power spectrum. As the dimensionality of the pseudo-expansion is increased, the weight of the low-energy modes grows and a noticeable tail emerges in the correlation function. Since the pion modes having the lowest energy are amplified the most, this effect is somewhat suppressed by the relative smallness of the box (the lowest quasi-particle momentum is given by $k_{\rm min} = L/2\pi$). This is clearly borne out by the results for a larger box, $L = 8$ fm, which display a more pronounced long-range enhancement.

It is interesting to note that the mean-field dynamics (solid curves) leads to a very good reproduction of the corresponding results obtained with the unapproximated equation of motion (solid symbols). This constitutes additional support for the utility of the approximation.

The correlation functions can be conveniently characterized by their full width at half maximum. For $L$=8 fm, this quantity is extracted as 1.0 (1.0),1.45 (1.30), 2.75 (2.60) using the exact (approximate) equations of motion with $D = 1, 2, 3$, respectively. These values reflect the sensitivity of the *DCC* phenomenon to the cooling rate.

## 4.3 Fraction: idealized sources

It has been proposed [2, 3, 4, 5] that the occurrence of disoriented chiral condensates may be revealed by observing the neutral pion fraction, $f = n_0/n$, where $n_0$ is the number of neutral pions observed and $n = n_- + n_0 + n_+$ is the total pion number. In the extreme idealized case where the pion field is uniquely oriented in a particular $O(3)$ direction (*i.e.* the pion field has that particular orientation throughout the source region), only pion modes with that particular isospin alignment can be produced. Since that particular direction of the pion field is entirely random, due to the isospin symmetry, it is then a matter of elementary geometry to derive the resulting probability density for $f$ [3, 4],

$$P_1(f) = \frac{1}{2\sqrt{f}} \; . \tag{42}$$

This result is shown in the first panel of fig. 3. It deviates qualitatively from the binomial form expected for a normal scenario in which the charge states of the emitted pions are determined essentially independently. Consequently, the distribution of the neutral pion fraction, $P(f)$, may provide a suitable signal for the *DCC* phenomenon. However, when assessing the prospects for a measurable signal to actually appear, it is important to consider less idealized scenarios, as we shall do below.

The simple inverse square-root behavior (42) occurs only when a single domain is formed within which the pion field is fully aligned. However, as a first step towards more realistic scenarios, is instructive to consider several different domains, each one having its own particular isospin orientation. We shall first consider $N$ such sources, assuming for simplicity that each contributes the same total number of pions, $n$.

When only two domains are considered, $N = 2$, the resulting probability density is easy to calculate by means of an elementary convolution, $P_2 = P_1 * P_1$,

$$P_2(f) = 2\int df'df''\; P_1(f')\; P_1(f'')\; \delta(f'+f''-2f) \;=\; \left\{ \begin{array}{ll} \frac{\pi}{2} & (f \leq \frac{1}{2}) \\ \frac{\pi}{2} - 2\theta & (f \geq \frac{1}{2}) \end{array} \right. \;, \tag{43}$$

where $\cos\theta = 1/\sqrt{2f}$ [24]. It is also shown in fig. 3. By performing successive convolutions, $P_N = P_{N-1} * P_1$, it is then straightforward to obtain $P(f)$ for larger values of $N$. The resulting probability densities are shown in fig. 3 for up to $N$=6 sources of equal size, beyond which the trend is clear and fairly uninteresting. The mean value of the neutral fraction remains equal to one third for all values of $N$, while its variance decreases steadily, $\sigma_f^2 = 4/45N$. The anomalous behavior for a small number of sources quickly reverts to a Poisson-like distribution as several sources contribute. These curves are the same as those recently presented in ref. [29]. We also note that the calculations made for fig. 3 in ref. [8] for two and three idealized sources agree well with the exact results (taking account of the Monte Carlo error).

It is interesting to see how the above result is modified when the emission has a more statistical character. For that purpose we assume that the pion multiplicity distribution has an exponential form, for each of the three charge states and for each of the $N$ sources independently. One might regard this modification as a simple means of modeling an ensemble of $N$ sources having different sizes (as measured by the multiplicity of pions contributed). Since the overall normalization of the multiplicity is immaterial for the present analysis, the (relative) multiplicity $\nu$ can then be considered as a continuous variable. Thus, the normalized $\pi_0$ multiplicity distribution from a single source would be $p_1(\nu_0) = \exp(-\nu_0)$ which has unit mean and variance. The normalized multiplicity distribution arising from $N$ similar sources can again be obtained by successive convolution leading to Poisson distributions of ever higher order, $p_{N+1}(\nu_0) = \nu_0^N \exp(-\nu_0)/N!$. In a given event, the neutral fraction is then $f = \nu_0/\nu$ where $\nu = \nu_- + \nu_0 + \nu_+$ is the total number of pions emitted from the $N$ sources in one particular event. It is elementary to show that the corresponding distribution is $P'_1(f) = 2 - 2f$. Furthermore, for two sources,

$$P'_2(f) \; = \; (P'_1 * P'_1)(f) \; = \; \begin{cases} \frac{32}{3}(\frac{3}{2}f - 3f^2 + f^3) & (f \le \frac{1}{2}) \\ \frac{32}{3}(1-f)^3 & (f \ge \frac{1}{2}) \end{cases} \; , \tag{44}$$

and increasingly complicated algebraic expressions hold for larger $N$. The mean value remains equal to one third, of course, and the associated variance is $\sigma_f^2 = 1/18N$.

The resulting distributions $P'_N(f)$ are included in fig. 3 (dashed curves). There is a remarkable similarity between the distributions obtained in the two different scenarios, although $P'(f)$ has a width that is smaller than that of $P(f)$ by a factor of $\sqrt{5/8} \approx 0.79$, consistent with the more statistical nature of the emission. Other idealized emission scenarios have been considered. In particular, Cohen *et al.* [30] considered the effect of quantum-mechanical isospin correlations, and very recently Anselm and Ryskin [31] calculated $P(f)$ for a rapidly varying source field. The various analyses lead to the general expectation that the neutral fraction can display an observable signal only if it arises from very few independent sources.

## 4.4 Fraction: dynamical simulations

After the above analysis of idealized cases, we consider now neutral fractions that have been generated by dynamical simulations of the type leading to the power spectra and correlation functions discussed above, namely pseudo-expansions of thermally sampled chiral fields in one, two, and three dimensions. The results are based on samples of 100 configurations in a cubic torus with $L = 5$ fm, the same as those employed for the calculation of the power spectra and the correlation functions.

The neutral pion fraction is sensitive to event-by-event fluctuations in the individual multiplicities and a proper treatment must therefore invoke the underlying quantum state represented by the given field configuration. The simplest manner of making such an association is to assume that the given pion field $\boldsymbol{\pi}(\boldsymbol{r}, t)$ represents a coherent many-body state [32], $|\boldsymbol{\pi}> \sim \exp(\sum_{\mathbf{k}} \boldsymbol{\pi}_{\mathbf{k}} \cdot \boldsymbol{a}^\dagger)|0>$. This form leads to a Poisson multiplicity distribution and, as a consequence, it contains arbitrarily large charge fluctuations, which appears to be somewhat unrealistic. More complicated associations displaying more realistic features have also been suggested (see ref. [27], for example) and they lead to various types of correlated emission. An instructive analysis of several suggested associations was given recently by Amado and Kogan [24]. In the present work, we assume for simplicity that the multiplicities are given by the corresponding expectation values, so that the number of neutral pions contributed by a particular mode $\mathbf{k}$ is $n^0_{\mathbf{k}} = < n^0_{\mathbf{k}} > \sim \omega_k |\pi^{(3)}_{\mathbf{k}}|^2$ and the corresponding total number of pions arising from that mode is $n_{\mathbf{k}} = < n_{\mathbf{k}} > \sim \omega_k \boldsymbol{\pi}^*_{\mathbf{k}} \cdot \boldsymbol{\pi}_{\mathbf{k}}$. The above analysis of idealized sources, sect. 4.3, suggests that the inclusion of incoherent multiplicity fluctuations would tend to degrade the *DCC* signal.

Let us first discuss the results obtained with the unapproximated equation of motion (1). The calculated distributions of the neutral fraction, $P(f)$ are displayed in fig. 4 (left panels). When extracting $f$ we have considered different classes of pions, corresponding to various idealized experimental cuts. The most restrictive cut admits only those pions that have zero momentum ($k = 0$), which arise from the isospin oscillations of the order parameter, $\underline{\boldsymbol{\pi}}(t)$. The resulting probability density for the neutral fraction (solid dots connected by solid lines) follows very well (to within the numerical sampling errors) the anomalous $1/\sqrt{f}$ form corresponding to the idealized one-source scenario (thin solid curve). This feature is a simple consequence of isospin symmetry and the fact that only the single mode having vanishing $k$ contributes.

The second class of pions consists of those having a kinetic energy $E$ below 200 MeV. This inclusion of the lowest-lying modes with finite momenta already significantly modifies the distribution of the neutral fraction and it now exhibits a broad peak with a maximum near one third. Since the number of pions arising from the oscillations of the order parameter is relatively small, as illustrated by the power spectra in fig. 1, they have little influence on the resulting neutral fraction, once even the lowest quasi-particle modes are included as well. This feature is illustrated by the similarity between the dashed and the dotted curves which were obtained by either including or excluding the zero-momentum pions, respectively. Although the resulting distribution is still anomalous due to its large width, it no longer displays the striking divergence near $f \approx 0$ and so it does not present a qualitative signal.

The third class of pions considered includes all the produced pions without regard to their energy. The associated neutral fraction distributions are relatively narrowly peaked, and more so the slower the cooling is. They are centered around one third and the widths resemble what would be expected for statistical production of the pions. For example, the distribution obtained for $D$=3 resembles what would be obtained with the statistical emission model considered above, using about nine similar sources. We also note that the present results are in accordance with earlier studies [8] in which a simplified dynamical equation was used to demonstrate that large systems ($L \approx 16$ fm) produce binomial distributions whereas sufficiently small systems ($L$=4 fm) display neutral fractions similar to those obtained for $N \approx 3$ idealized sources.

When considering the dependence of $P(f)$ on the dimensionality of the pseudo-expansion, we observe that the anomalous behavior grows more pronounced as $D$ increases. This is a consequence of the fact that the anomalous multiplicity distributions are primarily generated by the low-energy modes and they experience an increasing degree of amplification as the cooling rate is increased [18]. Even the distribution based on all the pions is noticeably wider for the fastest expansion scenario.

In order to assess the utility of the mean-field approximation as a tool for dynamical studies, we have treated the same three pseudo-expansion scenarios with the approximate dynamics and the resulting $f$ distributions are included in fig. 4 (right panels). The agreement between the two sets of results is remarkable. In fact, there appears to be correspondence within the errors arising from the finite sample size (100 evolving field configurations have been considered in each scenario). Therefore, for the extraction of this type of observable, the mean-field approximation may indeed be useful.

It is perhaps most striking that the results for all the different expansion scenarios are so similar. This feature suggests that the appearance of the neutral fraction distribution is largely decoupled from the profile of the power spectrum. Indeed, our simulations suggest that the distribution neutral pion fraction, $P(f)$, may exhibit an anomalous behavior even if the expansion is farily slow. However, as soon as pions with finite momenta are included, $P(f)$ has the appearance of a broad peak rather than a decreasing function, so it is qualitatively similar to the normal behavior and careful quantitative analysis is then called for in order to identify the *DCC* phenomenon. It should also be kept in mind that since the anomalous behavior of $P(f)$ is driven by the low-energy modes, the practical observability of the phenomenon depends on the abundance of low-energy pions. This inherent feature poses several problems. One is the simple geometrical fact that there are relatively few pion modes having small momenta, so the signal-carrying pions are relatively scarce. Another problem is that such pions are relatively difficult to detect because they move close to the beam. Moreover, especially in collider experiments, they may be below detection threshold. A rough indication of the relative multiplicity of low-energy pions is provided by the power spectra in fig. 1 and, consequently, the amplification produced by fast expansions are still expected to be helpful, perhaps even essential for the detectability of the *DCC* phenomenon. The effective dimensionality of the expansion (or more directly: the rate at which the field fluctuations subside) is thus a question of key importance to the prospects for observing disoriented chiral condensates in the laboratory.

### 4.4.1 Equalization

We present here a further analysis that serves to better illuminate why it is so relatively hard to produce real *DCC* "domains" during the dynamical evolution.

We first note that the initial field configurations tend to possess a large degree of isospin symmetry. This is especially evident when they are sampled in thermal equilibrium at high temperature, but it is expected to be generally the case, since there are yet no isospin-breaking mechanisms in play at that early time (if there were, then they should better be included in the *DCC* studies). The only deviations from perfect rotational invariance in isospace is associated with the statistical fluctuations in the population of the individual quasi-particle modes. It then follows that the $O(3)$ tensor describing the fluctuations perpendicular to the order parameter, $\prec \delta\boldsymbol{\phi}_\perp \delta\boldsymbol{\phi}_\perp \succ$, is approximately isotropic. Consequently, the same is true for the perpendicular part of the quasi-particle mass tensor $\boldsymbol{M}$.

It is then important to understand how such initial imperfections develop in the course of time, as the system expands and cools. It is only the lowest pion modes that may be amplified in the course of the pseudo-expansion [18], as is borne out by the resulting power spectra displayed in fig. 1. Since the contribution of the lowest modes to the field fluctuations is relatively small (even after any amplification they may have experienced), the feedback of such amplification onto the mass tensor is generally small. The mass tensor can then be regarded as approximately isotropic throughout and the different isospin components are then amplified about equally. There is thus no tendency for "domain growth", *i.e.* a catastrophic isospin alignment of different quasi-pion modes. In fact, any isospin imbalance in the mass tensor will quickly be counteracted by the amplification (see below): If a given isospin direction has larger-than-average fluctuations, the associated effective mass squared is then also larger than average, and the amplification rate is correspondingly reduced, thereby bringing the fluctuation tensor closer to isotropy. Thus, any initial anisotropy will at best be preserved, but is generally degraded by the dynamical effect of the self-interaction.

In order to further elucidate this equalizing mechanism, one may analyze an extreme (and unrealistic) scenario in which the dominant pion strength is carried by unstable modes. It is then elementary to show that any anisotropy in the fluctuations would quickly decay. To appreciate this important feature, it suffices to consider a simpler world in which the pion field is an $O(2)$ vector, $\boldsymbol{\pi} = (\pi_-, \pi_+)$. In this schematic model, let the initial fluctuations be given by $< \delta\pi_\pm^2 >= \frac{1}{2} < \delta\pi^2 > (1 \pm \epsilon)$, where the small parameter $\epsilon$ measures the anisotropy in the isoplane. The corresponding effective quasi-particle masses $\mu_\pm$ are then given by

$$\mu_\pm^2 = \lambda[< \sigma^2 > + < \delta\pi^2 > (2 \pm \epsilon) - v^2] \; = \; \mu_0^2 \; \pm \; \epsilon\lambda < \delta\pi^2 > , \tag{45}$$

and the dispersion relations are $\omega_\pm^2 = \omega_k^2 \pm \epsilon\lambda < \delta\pi^2 >$, where $\omega_k^2 = k^2 + \mu_0^2$ with $\mu_0$ being the effective mass for $\epsilon$=0. Inside the region of instability (where $\omega_\pm^2 < 0$) the fluctuations are then amplified as $< \delta\pi_\pm^2 > \sim \exp(2|\omega_\pm|t)$. Consequently, the anisotropy decays approximately exponentially,

$$\epsilon(t) \; \equiv \; \frac{< \delta\pi_+^2 > - < \delta\pi_-^2 >}{< \delta\pi_+^2 > + < \delta\pi_-^2 >} \; \approx \; \epsilon(0) \; \mathrm{e}^{-\frac{2\lambda}{|\omega_k|} < \delta\pi^2 > t} \; . \tag{46}$$

The associated characteristic rate is $t_k^{-1} \approx 2\lambda < \delta\pi^2 > /|\omega_k|$ for a given magnitude of the wave number. Inserting numbers that are typical of realistic scenarios, we find that the decay time is only a fraction of one fm/$c$. Since this time is relatively short in comparison with the characteristic duration of the amplification process (which is 1-2 fm/$c$ [18]), the dynamics will tend to suppress rather than enhance any anisotropy in the isospin distribution of the quasi-pions.

# 5 Concluding discussion

The possible formation of disoriented chiral condensates in high-energy hadron and heavy-ion collisions has generated considerable interest in the past few years, since the associated anomalous pion multiplicity distributions may offer a means for testing our understanding of chiral symmetry. In order to elucidate the conditions for the occurrence of *DCC* phenomena and the prospects for their experimental detection, it is necessary to carry out dynamical simulations of the non-equilibrium evolution experienced by the chiral field as it relaxes from an initially very excited state, in which chiral symmetry is approximately restored, towards the normal vaccuum in which the symmetry is significantly broken.

The most popular tool for such dynamical studies has been the linear $\sigma$ model in which the chiral degrees of freedom are described by an $O(4)$ real classical field subject to a simple non-linear interaction. Even though this description is relatively simple, ignoring all other degrees of freedom (such as those represented by other meson type or individual quarks), it still presents a significant computational challenge. It is therefore of practical interest to develop useful approximate methods for solving the non-linear equation of motion for the chiral fields.

In the present paper we have presented a simple mean-field approximation that is based on the Hartree factorization technique and retains all the terms in a $1/N$ expansion. It describes the system in terms of an evolving order parameter and a gas of independent quasi-particles whose interaction is via their self-consistent medium-modified effective mass tensor. This can be regarded as a dynamical extension of the treatment developed in ref. [21] in the context of statistical equilibrium and it presents a simple and useful conceptual framework for analyzing the complicated *DCC* dynamics, as already illustrated in ref. [18]. Because of the effective decoupling, the mean-field dynamics can be solved very efficiently in the $k$ representation, leading to a savings in computational effort by over an order of magnitude.[2]

The present treatment thus relies on a succession of approximations which can be briefly summarized as follows. The general framework is the familiar linear $\sigma$ model in its classical form where real $\sigma$ and $\boldsymbol{\pi}$ fields experience an interaction containing up to the fourth power of the fields. Furthermore, the fields are subjected to periodic boundary conditions and their initial form is chosen to reflect Bose-Einstein statistics (see below for a further discussion). On top of these general features, the present work develops a "mean-field" approximation in which the quasi-particle degrees of freedom interact only via their self-consistent effective masses.

It is instructive to note the analogy of the present treatment with the semi-classical mean-field models that have proven very successful in the context of nuclear reaction dynamics at lower energies. The closest analogue is the Vlasov approximation in which the individual nucleons move independently in a common self-consistent mean field (as the motion of the quasi-particles in the present treatment is governed by the self-consistent mass tensor). In order to take account of their underlying quantal nature, the initial phase space distribution is specified in accordance with the Fermi-Dirac statistics governing real nucleons (as the present treatment employs the Bose-Einstein statistics appropriate to the quasi-particle meson gas). This type of approach is justified by the fact that the time scales associated with the inevitable reversion to Maxwell-Boltzmann statistics are long in comparison with the typical durations of the processes under study (whether nuclear reactions or *DCC* formation).

In addition to presenting the formal development of the mean-field treatment, we have also sought to illustrate its practical utility by testing its accuracy for a number of observables of particular interest in the quest for observing *DCC* phenomena. The comparisons with the corresponding results of the unapproximated non-linear equation of motion suggest that the developed mean-field approximation is indeed quantitatively useful for this kind of simulation.

[2]For the unapproximated equation of motion the non-linear character of the interaction renders the $k$ representation inefficient and the $x$ representation is preferable, even though the Fourier transformations associated with the initialization and the spectral analyses are relatively costly. The computational effort is then roughly proportional to the number of lattice points, $(L/dx)^3$, where the lattice spacing is $dx \approx 0.2$ fm. In the mean-field approximation, there is no direct coupling between the different modes and the $k$ representation is then the economical choice. The associated effort is proportional to the number of modes included, $(2K+1)^3$, where $L/K \approx 1$ fm is the minimum wave length considered. The resulting gain factor is then $\sim (5/2)^3$ and exceeds ten.

The calculated results are interesting in their own right, as they illustrate various key features of the *DCC* phenomenon. We employed a cubic volume with a side length of $L = 5$ fm as representative of source sizes expected in heavy-ion collisions. A larger source would behave similarly in the present idealized kind of calculation. In particular, since the *DCC* anomaly is associated primarily with the low-energy pions (*i.e.* the signal-carrying pions have close to zero momentum in the frame of the source), the ideal $1/\sqrt{f}$ behavior of $P(f)$ will appear only if the observation is restricted to $k = 0$. Consequently, as the box is enlarged this single component will constitute an ever smaller fraction of the total power. This feature illustrates how the collection of pions from a large spatial volume tends to wash out the *DCC* signal. A cleaner signal emerges for a smaller volume, so a key question is how small phase-space domains can be experimentally separated.

Moreover, we have concentrated on evolutions prepared at $T_0 = 400$ MeV, in order to increase the chance for the system to enter the unstable regime where exponential amplification occurs [18]. If the initial temperature were lower, the magnification of the low-lying pion modes would be smaller as well, or entirely absent, but the results for the pion fraction would not change appreciably, as is borne out by the insensitivity of $P(f)$ to the value of $D$.

Once the system has cooled sufficiently, the different modes effectively decouple. Each mode $\mathbf{k}$ contributes pions having an independent orientation in isospace, $\hat{\boldsymbol{\pi}}_{\mathbf{k}}$. The resulting pion multiplicity distribution can then be understood as resulting from a collection of such independent sources, in a scenario similar to that discussed in sect. 4.3. Accordingly, the resulting probability density $P(f)$ is the corresponding convolution of all the contributions from all the individual modes, with proper account taken of their different "sizes" as measured by the contribution to the total multiplicity. As an instructive aid to the interpretation of the dynamically calculated distributions of the neutral pion fraction, $P(f)$, we extended various earlier discussions of idealized emission scenarios, considering a number of similar but independent pion sources each contributing either a fixed or a stochastic multiplicity. This analysis brings out the difficulty in obtaining the ideal $1/\sqrt{f}$ signal in realistic scenarios when there are several different sources of the observed pions.

Simulations including more realistic features are clearly called for. One important limitation of simulations done with the standard linear $\sigma$ model is the absence of the many other degrees of freedom in the system. It is of interest to understand their effect on the *DCC* phenomenon. Moreover, the excited systems produced in high-energy collisions are embedded in the normal vacuum and they are born in a state of rapid expansion. Since this latter feature may be essential for the occurrence of *DCC* phenomena [18], it appears desirable to incorporate realistic initial expansion patterns. These must be obtained on the basis of more fundamental descriptions, such as parton dynamics, and until such information has become available the simulations of *DCC* phenomena can only be suggestive.

The author is pleased to acknowledge helpful discussions with many colleagues, includ-ing especially S. Gavin, J.I. Kapusta, Y. Kluger, V. Koch, E. Mottola, K. Rajagopal, R. Vogt, and X.N. Wang. This work was supported in part by the Director, Office of Energy Research, Office of High Energy and Nuclear Physics, Nuclear Physics Division of the U.S. Department of Energy under Contract No. DE-AC03-76SF00098.

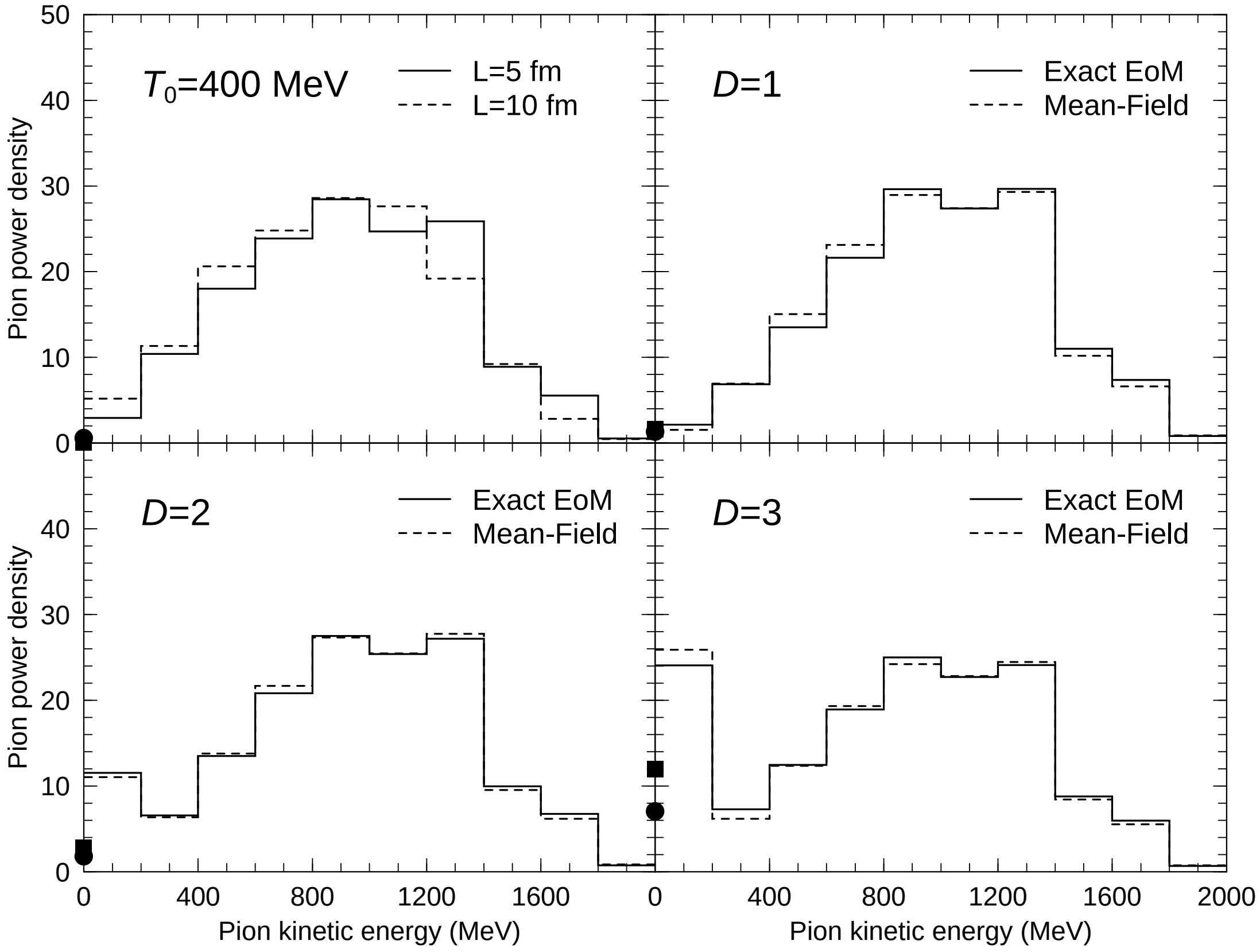

Figure 1: Pion power spectrum.

The upper-left panel shows the power spectrum of the transverse quasi-particle modes for thermal equilibrium at the temperature $T_0 = 400$ MeV. The system is enclosed in a cubic torus with size $L = 5$ fm (solid histogram); for comparison is also shown the results for $L = 10$ fm (dashed) (the irregularities are caused by the shell structure of the cube). The other three panels show the power spectrum of the pions emerging asymptotically as a result of propagating 100 field configurations that were sampled from the thermal distribution at $T_0$. The solid histograms were obtained by employing the exact equation of motion (1) while the dashed histograms are for the mean-field approximation, eqs. (9-11), in addition to the Rayleigh cooling term $-(D/t)\partial_t \boldsymbol{\phi}$ which roughly emulates a Bjorken-like expansion of matter in $D$ dimensions (ordinary static equilibrium is thus recovered for $D$=0). The normalization is such that the total power is the same in all cases. The power carried by mode having $k = 0$ (the order parameter) is indicated by the solid circle and square for the exact and approximate dynamics, respectively.

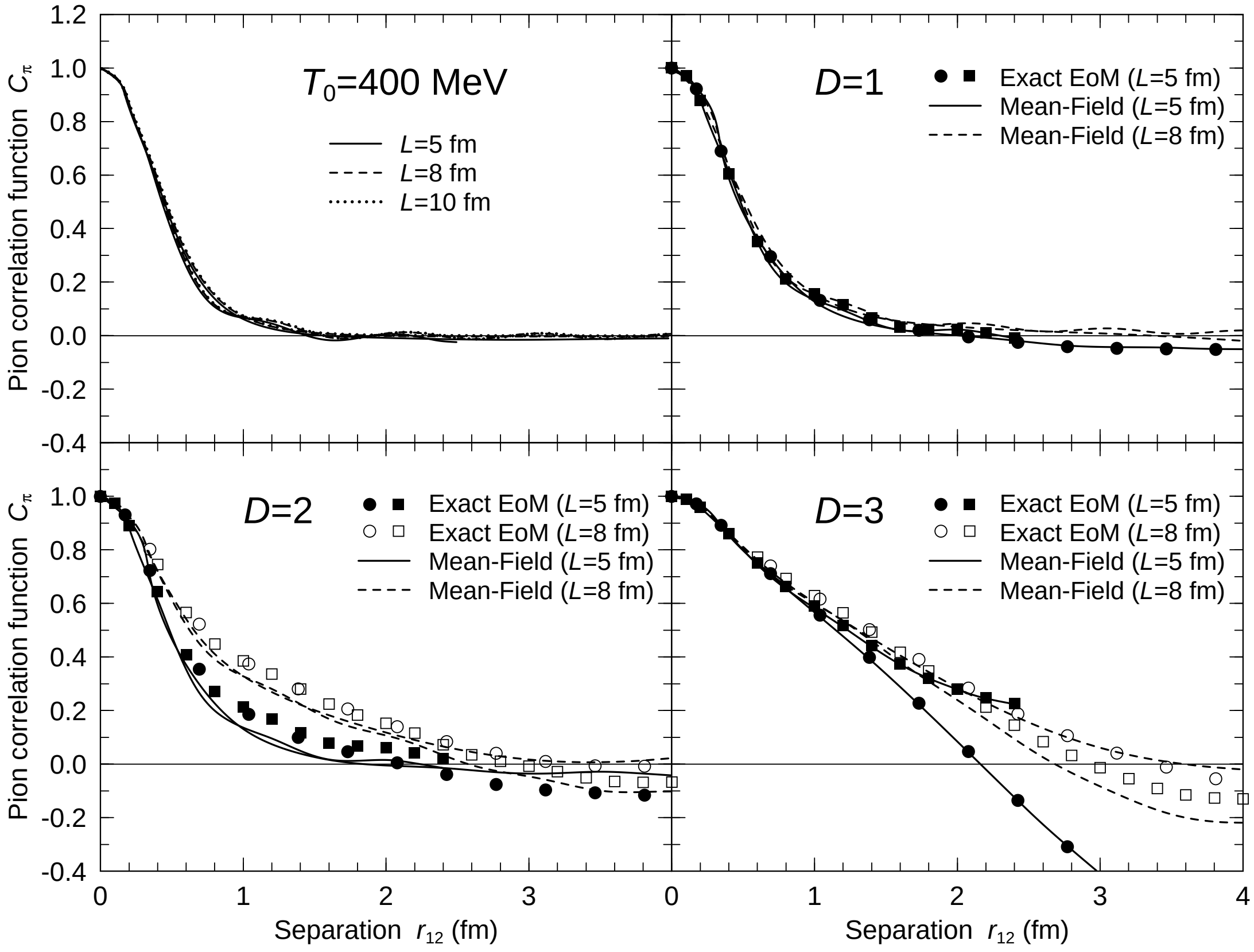

Figure 2: Pion correlation function.

The pion correlation function $C_\pi(r_{12})$ defined in eq. (41). The upper-left panel is the correlation function for the transverse quasi-particle modes in thermal equilibrium at the temperature $T_0 = 400$ MeV, obtained with various values of the box size, $L$=5,8,10 fm. The other three panels were obtained at large times after 100 field configurations were sampled from that thermal ensemble and subjected to a pseudo-expansion in $D$=1,2,3 dimensions, using either the exact equation of motion (1) (solid curves) or the mean-field approximation, eqs. (9-11) (dashed curves). The results obtained with the exact dynamics for a box with $L$=5 fm are shown by the solid symbols, as obtained by directing the relative separation $\boldsymbol{r}_{12}$ either along the lattice (squares) or diagonally (circles). Exact results obtained with $L$=8 fm are indicated by the corresponding open symbols. The results of the corresponding approximate dynamics are shown by the solid and dashed curves for $L$=5 fm and $L$=8 fm, respectively.

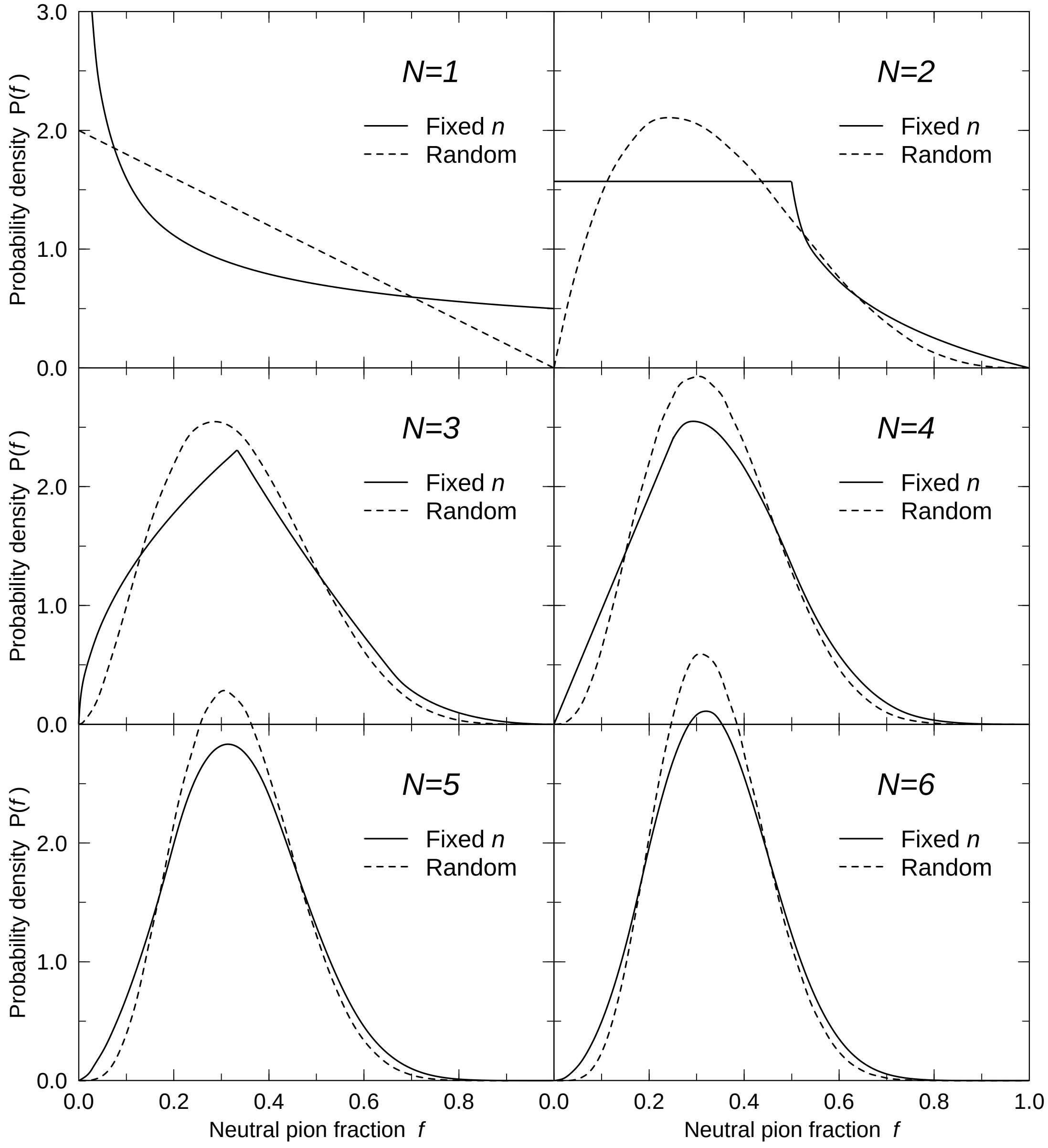

Figure 3: Neutral fraction from idealized sources.
The normalized probability density, $P(f)$, for the neutral fraction $f = n_0/n$ obtained for $N$ similar sources, assuming either the exact same multiplicity contribution $n$ from each source (solid curves) or an exponential distribution of the individual multiplicities $n_-$, $n_0$, $n_+$ for each source separately (dashed curves).

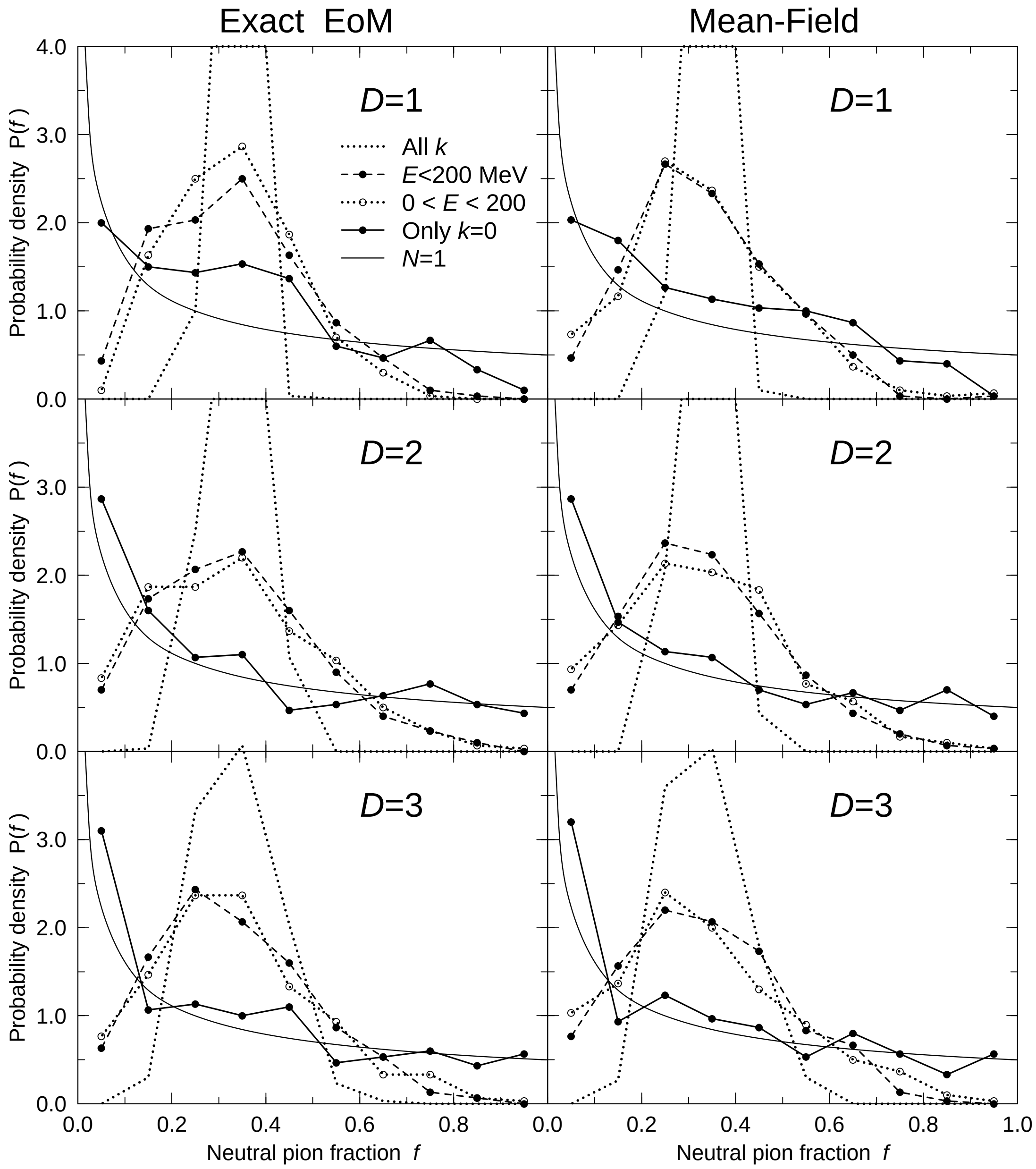

Figure 4: Neutral fraction from dynamical simulations.
The normalized probability density, $P(f)$, for the neutral fraction $f = n_0/n$ obtained by subjecting a thermal sample of 100 field configurations to a pseudo-expansion in $D$ dimensions, using either the exact equation of motion (1) (left) or the mean-field approximation (right). The various curves were obtained by considering different classes of pions when extracting $f$: those pions having $k = 0$ only (dots connected by solid lines), those with kinetic energies $E$ below 200 MeV either excluding or including $k = 0$ (dots connected by dotted or dashed lines, respectively), and all of the pions (dotted curve). For reference is shown the result for the idealized case of a single source (light solid curve), also shown in the upper-left panel in fig. 3.